\documentclass[pra,reprint,floatfix,showpacs,superscriptaddress]{revtex4-1}
\usepackage{amssymb,amsmath}
\usepackage{graphicx}
\usepackage{subfigure}
\usepackage{bm}
\usepackage[varg]{txfonts}
\usepackage{color}

\def\be{\begin{equation}}
\def\ee{\end{equation}}
\def\bea{\begin{eqnarray}}
\def\eea{\end{eqnarray}}

\def\F1k{\widehat F_1({\bm k}^\prime)}

\def\tk1k{\widehat T_{{\bm k}_1,{\bm k}^\prime}}

\def\g1k{\widehat G^{-1}_{\bm k}}

\def\kk1{{{\bm k},{\bm k}_1}}

\def\bk{ \mathbf{k} }
\def\bq{ \mathbf{q} }
\def\br{ \mathbf{r} }
\def\bmatrix{\begin{pmatrix}}
\def\ematrix{\end{pmatrix}}

\begin{document}

\title{
Topological Uniform Superfluid and FFLO Phases in 3D to 1D crossover of spin-orbit coupled Fermi gases}

\author{Kangjun Seo}
\affiliation{Department of Physics and Astronomy, Clemson University, Clemson, SC 29634, USA}
\author{Chuanwei Zhang}
\affiliation{Department of Physics, The University of Texas at Dallas, Richardson, TX 75080 USA}
\author{Sumanta Tewari}
\affiliation{Department of Physics and Astronomy, Clemson University, Clemson, SC 29634, USA}

\begin{abstract}
We consider the quasi-one dimensional system realized by an array of weakly coupled parallel one-dimensional ``tubes'' in a two-dimensional lattice which permits free motion of atoms in an axial direction in the presence of a Zeeman field, Rashba type spin orbit coupling (SOC), and an $s$-wave attractive interaction, while the radial motion is tightly confined. We solve the zero-temperature ($T=0$) Bogoliubov-de Gennes (BdG) equations for the quasi-1D Fermi gas with the dispersion modified by tunneling between the tubes, and show that the $T=0$ phase diagram hosts the Fulde-Ferrell-Larkin-Ovchinnikov (FFLO) phase with non-zero center of mass momentum Cooper pairs for small values of the SOC while for larger values of the SOC and high Zeeman fields the uniform superfluid phase with zero center of mass momentum Cooper pairs has an instability towards the topological uniform superfluid phase with Majorana fermions at the tube ends. Also, we show that tuning the two-dimensional optical lattice strength in this model allows one to explore the crossover behaviors of the phases during the transition between the 3D and the 1D system and in general the FFLO (for small SOC) and the topological uniform superfluid phase (for large SOC) are favored as the system becomes more one-dimensional. We also find evidence of the existence of a Zeeman tuned topological quantum phase transition (TQPT) within the FFLO phase itself and for large values of the Zeeman field and small SOC the TQPT gives rise to a topologically distinct FFLO phase.\end{abstract}

\maketitle


\section{Introduction}
Since the Fulde-Ferrell-Larkin-Ovchinnikov (FFLO) superconducting/superfluid phase, which is a condensate of Cooper pairs with finite center of mass momentum and the spatially modulated order parameter~\cite{fulde-1964,larkin-1964}, was proposed in the 1960s, the theoretical and experimental studies have been conducted in a variety of areas, ranging from heavy fermions~\cite{bianchi-2003}, dense quark matter~\cite{alford-2001}, ultracold atomic gases~\cite{mizushima-2005,sheehy-2006,casal-2004,parish-2007}, and so forth.
This unconventional pairing state can be realized by pairing the particles on different Fermi surfaces that are mismatched in spin-polarized systems~\cite{koponen-2007}, as well as by pairing the particles on a single Fermi surface that is deformed by breaking the spatial inversion symmetry~\cite{zeng-2013}.
However, the region occupied by the FFLO phase in the $T=0$ phase diagram of ultracold fermions is typically small in 2 and 3 dimensions and becomes diminished with increasing temperature~\cite{hu-2006,bulgac-2008}. In the 1D ultracold atomic Fermi gas with spin imbalance, on the other hand, the FFLO phase occupies a much larger portion of the phase diagram \cite{liu-2007,liao-2010}, although both the FFLO phase (with spin imbalance) and the conventional Bardeen-Cooper-Schrieffer (BCS) superfluid phase (with equal number of up and down spins) are only power law ordered in 1D even at $T=0$.
Starting from the Bethe ansatz equations~\cite{orso-2007,hu-2007}, one finds that the 1D polarized Fermi gas can host an unpolarized uniform superfluid phase, an FFLO superfluid phase, a fully-polarized normal phase, and the vacuum.

Although in 1D the FFLO phase occupies a large region of the $T=0$ phase diagram, the study of 1D system in the frame of mean-field theory is problematic. This is because the pair fluctuations become significant with the reduced dimensionality and true long-range order in 1D is completely destroyed by the Mermin-Wagner Theorem but has only power-law correlations in the thermodynamic limit~\cite{yang-2001}. Also, experimental investigations of the superfluid phases are hard since the transition temperature $T_c$ of the 1D superfluid is zero.
In order to overcome these technical difficulties peculiar to pure 1D systems,
the two-dimensional array of 1D atomic ``tubes''~\cite{moritz-2005} has been proposed as a model of quasi-1D Fermi gas, where atoms are allowed to move freely in one direction and tightly confined in the transverse motion~\cite{parish-2007}. In this weakly coupled 1D system the atomic motion is gas-like in the axial direction but the weak coupling in the transverse lattice plane stabilizes the long-range superfluid order. The stabilization of the long-range order allows the use of mean field theory while the anisotropy between the couplings in the axial direction and the transverse plane still reveals the behavior specific to 1D.
In this model, the mean-field study shows that the parameter space of the FFLO superfluid phase is enhanced in a spin-polarized quasi-1D Fermi gas as the confinement of the atomic motion in the transverse directions is enhanced (that is, as the system becomes more and more 1D-like).
In addition, this model allows one to investigate the behavior of the system during the crossover from the 3D to 1D regimes by increasing the optical lattice strength that controls the transverse hopping between the tubes.

The presence of spin-orbit coupling (SOC) and external Zeeman fields in otherwise uniform $s$-wave superfluids/superconductors is another topic of great current interest. These effects in combination with long-ranged superfluid order can produce interesting phases such as non-centrosymmetric superconductors~\cite{agterberg-2003,samokhin-2004,kaur-2005} and nonuniform superfluids in ultracold systems~\cite{zeng-2013,wu-2013,liu-2013}.
In addition to these rich quantum phases, the combination of SOC and a suitably directed Zeeman field allows one to take a further step towards novel quantum states, the topological superconducting/superfluid phases~\cite{zhang-2008,jiang-2011,gong-2012,seo-2012,seo-2012-1,seo-2013,qu-2013,qu-2013-07,zhang-2013}.
In solid state systems, spin-orbit coupled semiconducting thin films and nanowires with proximity induced $s$-wave superconductivity and a Zeeman field can host novel non-Abelian topological states with the order parameter defects such as the vortex cores and the sample edges supporting localized topological zero-energy excitations called Majorana fermions ~\cite{Sau,Long-PRB,Roman,Oreg,mao-2012}. Since the SOC and Zeeman field in $s$-wave superfluids induce topological superfluid phases in 1D with Majorana fermion end states, while at the same time the crossover from 3D to quasi-1D in the presence of a Zeeman field enhances the stability of the FFLO phase, it is a natural question to ask what is the precise behavior of the system as it crosses over from 3D to 1D but in the presence of both spin-orbit coupling and a Zeeman field. In particular, we ask which of the two phases between the (a) topological uniform superfluid (that contains the Majorana fermion end states and an otherwise gapped spectrum) and (b) the FFLO non-uniform superfluid (that contains non-zero center of mass momentum Cooper pairs and may be gapless) is preferred as the system crosses over from 3D to quasi-1D in the presence of both Zeeman field and SOC.
In this paper we address this question by calculating the $T=0$ phase diagram of a set of parallel 1D ``tubes'' arranged in a 2D lattice in the presence of a Zeeman field and an additional Rashba type spin-orbit coupling.

\begin{figure}[t]
\includegraphics[width = 0.7\linewidth]{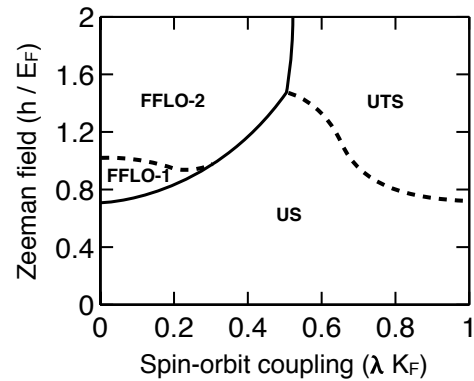}
\caption{
The zero-temperature ($T$=0) mean-field phase diagram in the ($h,\lambda$) plane for quasi-1D Fermi gas computed using dimensionless interaction parameter $\gamma = -2$ and transverse hopping $t/ E_F = 0.01$.
The phase diagram hosts the uniform non-topological superfluid (\textit{US}), the uniform topological superfluid (\textit{UTS}), and the non-uniform FFLO phases (\textit{FFLO-1}, \textit{FFLO-2}) with the single-plane-wave Fulde-Ferrell (FF) approximation.
The solid line ($h_c$) represents the phase boundary between the uniform ($\bq = 0$) and the non-uniform ($\bq \neq 0$) superfluid phases.
The \textit{FFLO-1}--\textit{FFLO-2} and \textit{US}--\textit{UTS} transitions across the dashed line ($h_t$) are the topological quantum phase transitions (TQPT).
Note that application of higher Zeeman field ($h/E_F > 2$) results in the phase transition from the FFLO phase to fully polarized normal state only at $\lambda = 0$.
}
\label{fig-1}
\end{figure}

In this paper, we consider the mean-field theory of the above quasi-1D Fermi gas under the effective Zeeman field together with synthetic spin-orbit coupling, and
show that the $T=0$ phase diagram can host both the uniform superfluid state with zero center of mass momentum Cooper pairs, the uniform topological superfluid with Majorana fermions edge states, and a FFLO state with non-zero center of mass momentum Cooper pairs. Most importantly, we find that the uniform topological superfluid and the FFLO state occur in distinctly different regimes of the (Zeeman field - SOC) phase diagram (see Fig.~\ref{fig-1})). From the uniform (non-topological) superfluid state the FFLO state is the leading instability at high enough Zeeman fields but only for small values of the spin-orbit coupling. For larger values of the SOC, the FFLO phase disappears from the phase diagram and the uniform non-topological superfluid state transitions into a uniform \textit{topological} superfluid state for high values of the Zeeman field. This result will be experimentally very useful because it clearly separates the two interesting unconventional superfluid phases -- topological superfluid and FFLO -- into distinct regimes of the phase diagram as the system crosses over from 3D to 1D with the tuning of the transverse lattice.

\begin{figure}[t]
\includegraphics[width = 1.0\linewidth]{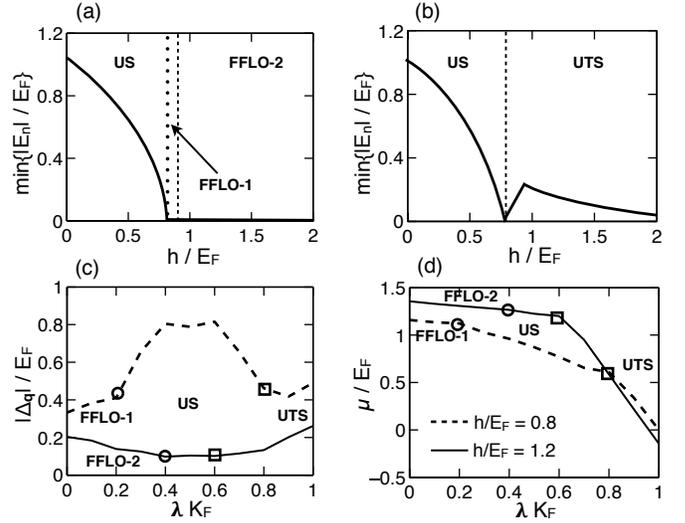}
\caption{
Zeeman field $h$ dependence of the energy gap (top panels) and the order parameter $\Delta_\bq$ and chemical potential $\mu$ as a function of $\lambda$ (bottom panels).
(a) The energy gap for a given SOC $\lambda K_F = 0.2$, showing the \textit{US}, and the gapless FFLO phases with increasing $h$. The dotted line indicates the phase boundary $h_c$, while the dashed line indicates the topological phase boundary $h_t$.
(b) The energy gap for the large SOC $\lambda K_F = 0.8$.  With increasing $h$, the gap closes at $h = h_t$, indicating the TQPT from the \textit{US} to the uniform topological superfluid phase (\textit{UTS}). The gap in the bulk spectrum protects zero energy Majorana fermion states at the ends of the 1D tubes.
(c) The order parameter $\Delta_\bq$ and (d) the chemical potential $\mu$ as a function of spin-orbit coupling $\lambda$ for different values of Zeeman fields $h/E_F = 0.8$ (dashed line) and $h/E_F = 1.2$ (solid line). Open circle and square mark the \textit{US} -- FFLO phase boundary ($h_c$) and the \textit{US} -- \textit{UTS} phase boundary ($h_t$), respectively.
}
\label{fig-2}
\end{figure}

The emergence of the FFLO phase with the Zeeman field for small values of the spin-orbit coupling is easily understood from the fact that the dominant instability of the system as it crosses over to 1D in the presence of a Zeeman field (but no SOC) is to the FFLO phase. For small SOC this instability persists and the system moves into the FFLO phase beyond a critical value of the Zeeman field via a first order phase transition. The SOC disfavors the FFLO phase, however, since it creates a mixing of the spins in the same band, thus increasing the tendency of the system to create spin-singlet $s$-wave pairs from the same band (note that, for symmetric Fermi surfaces, the FFLO phase is a result of pairing between electrons from the different spin-split bands). With increasing values of the SOC the system transitions from the FFLO phase to the uniform superfluid phase (in Fig.~\ref{fig-1} moving horizontally for a high enough Zeeman field) finally entering into the topological superfluid phase with Majorana fermions at a second order topological quantum phase transition (TQPT). We find that the FFLO phases are gapless (see FIG.~\ref{fig-2}(a)), while the uniform topological superfluid phase is gapped (see FIG.~\ref{fig-2}(b)) in the bulk but with zero energy Majorana fermion excitations at the ends of the 1D tubes.

In addition to these interesting phases, within the FFLO phase itself we find a topologically distinct FFLO phase (indicated as FFLO-2 in Fig.~1) setting in at higher values of the Zeeman field. This new FFLO phase is characterized by a non-trivial value of a certain relevant topological invariant known as Pfaffian invariant for 1D systems that usually signals, in the presence of a gapped spectrum, the emergence of a topological phase with Majorana fermion edge states. In contrast to the uniform topological superfluid state (shown as UTS in Fig.~1) with Majorana fermions end states, however, we find that the spectrum in both topologically distinct FFLO phases (FFLO-1 and FFLO-2) is gapless. A full characterization of these topologically distinct FFLO states is beyond the scope of the present paper and will be taken up in a future publication.

\section{Model}
The system we consider in this paper is an array of quasi-1D Fermi gas with two hyperfine states, $s =  \uparrow$, $\downarrow$, confined by a wide trapping potential $V_T(\mathrm{r})$. To create  the two-dimensional lattice of 1D tubes, we include optical lattice potential, $V_O(\mathrm{r}) = -U(\br) [ \cos (2\pi x /a) + \cos (2 \pi y / a)]$, with lattice spacing $a$ and depth $U(\br)$.
The latter breaks the 3D cloud apart into an array of parallel tubes aligned along one direction. We choose the axis of a tube as $z$ direction.
This model of two-dimensional array of 1D tubes has been studied for the FFLO superfluid phases in the quasi-1D spin-polarized Fermi gas.
With sufficiently low density and strong enough $U$, the particle motion along the tubes ($z$ direction) is well described by a parabolic dispersion with a single-band tight-binding dispersion for the $xy$ motion~\cite{parish-2007}.

In this paper, we will ignore the trapping potential $V_T(\br) = 0$ and discuss the spatially homogeneous quasi-1D system to simplify the analysis.
The spatial dependence of the physical quantities in a smooth trapping potential can be obtained by the local density approximation (LDA) by replacing the chemical potential of each spin state $\mu_s \to \mu_s(\br) = \mu_s - V_T(\br)$, while the difference of the chemical potential $\delta\mu(\br) = \delta\mu =  (\mu_\uparrow - \mu_\downarrow)/2$ to be held spatially constant.
Without the trapping potential $V_T(\br) = 0$, we consider the two-dimensional optical lattice with constant lattice depth $U(\br) = U$ comprised of $N_x \times N_y$ tubes, each of which has length $L_z$ along the $z$ direction.

Within this model, we include the Rashba type of spin-orbit coupling along the $z$ direction $\lambda k_z \sigma_y$ and the Zeeman splitting $h \sigma_z$, where $\sigma_z$ and $\sigma_y$ are the Pauli matrices. Then, we write the single particle Hamiltonian of spin-orbit coupled quasi-1D Fermi gas as
\be
H_0
=
\sum_{\bk,s,s^\prime}
\left[ (\epsilon_\bk -\mu_s) \delta_{ss^\prime} + \lambda k_z \sigma^{ss^\prime}_y - h \sigma_z^{ss^\prime} \right] c_{\bk s}^\dagger c_{\bk s^\prime},
\ee
where
$
\epsilon_\bk
=
\frac{k_z^2}{2m}
+
2 t
\left[
2 - \cos (k_x a) - \cos (k_y a)
\right].$
Here, $t$ is the transverse hopping between the nearest neighbor tubes, and $m$ is the atomic mass with letting $\hbar = 1$. $\lambda$ and $h$ are the spin-orbit coupling and Zeeman splitting strength, respectively.
$k_z$ is unconstrained in the thermodynamic limit, while $k_x$ and $k_y$ is in the Brillouin zone: $|k_x| \le \pi / a$ and $|k_y| \le \pi / a$.
Thus, $H_0$ describes the single particle's one-dimensional motion along the axial direction ($z$ direction) with spin states coupled to both SOC and Zeeman field, while radial motion ($xy$ motion) is restricted but can be controlled by adjusting transverse hopping $t$, permitting the crossover from 1D to 3D regime with increasing $t$.

With a broad Feshbach resonance, the interactions of highly dilute gases can be modeled by a contact interaction, which can be well described using a contact potential $g_{1D}\delta(z)$, leading to a many-body Hamiltonian
\be
H
=
H_0
+
\frac{g_{1D}}{L_z N_x N_y}
\sum_{\bk,\bk^\prime,\bq}
c_{\bq/2 + \bk \uparrow}^\dagger c_{\bq/2 - \bk^\prime \downarrow}^\dagger
c_{\bq/2 - \bk^\prime \downarrow} c_{\bq/2 + \bk \uparrow},
\ee
where $c_{\bk s}$ ($c_{\bk s}^\dagger$) is an annihilation (creation) operator of a fermion with momentum $\bk$ and hyperfine state $s$, and wavevector $\bq$.
In such a quasi-1D system, the scattering properties of the atoms in a single harmonic tube enable us to express $g_{1D}$ by the 3D scattering length $a_{3D}$,
$
1/g_{1D}
=
-m a_{1D}/2 = m a_\perp/2
\left( a_\perp/a_{3D} - A \right),
$
where $a_\perp$ is the characteristic oscillator length in the transverse direction, and the constant $A = -\zeta(1/2)/\sqrt{2} \sim 1.0326$ is enough for the confinement induced Feshbach resonance~\cite{bergman-2003,astra-2004}.
The attractive interaction in 1D gas can be achieved when $a_\perp / a_{3D} < A$.
For convenenience, we introduce the dimensionless interaction parameter for a homogeneous Fermi gas via $\gamma = -m g_{1D} / n = 2/n a_{1D}$, where $n = N /V$ is atomic density associated with the Fermi energy $E_F = K_F^2 / 2m$ with the Fermi wavevector $K_F = \pi n a^2$.
Thus, we can tune the 1D interaction via controlling the transverse size of the tubes.
$\gamma \ll 1$ corresponds to the weakly interacting limit, while the strong coupling regime is realized when $\gamma \gg 1$.

To study the 1D superfluidity in the presence of SOC and Zeeman splitting, we calculate the mean-field Bogoliubov–de Gennes (BdG) equation,
\be
H_{BdG}(z) \Psi_n (z) = E_n \Psi_n (z),
\ee
where $\Psi_n (z) = [ u_{\uparrow n}(z), u_{\downarrow n} (z), v_{\uparrow n} (z), v_{\downarrow n}(z) ]^{T}$ in the Nambu spinor representation and the BdG Hamiltonian $H_{BdG}$ is given by
\be
H_{BdG}(z)
=
\bmatrix
H_0(z) - h & -\lambda \partial_z & 0 & -\Delta(z) \\
\lambda \partial_z & H_0(z) + h & \Delta(z) & 0 \\
0 & \Delta^\ast (z) & -H_0(z) + h & \lambda \partial_z \\
-\Delta^\ast (z) & 0 & -\lambda \partial_z & -H_0(z) - h
\ematrix,
\ee
where $H_0(z) = -\frac{\partial_z^2}{2m} + 2t[2 - \cos (k_x a) - \cos (k_y a) ] - \mu$ and
\be
\label{eq-op}
\Delta (z) = -\frac{g_{1D}}{2N_x N_y} \sum_n [u_{\uparrow n} v_{\downarrow n}^\ast f(E_n) + u_{\downarrow n} v^\ast_{\uparrow n} f(-E_n)]
\ee
 is the order parameter.
Here, $f(x) = 1/ [e^{x/T} + 1]$ is the Fermi distribution function at temerature $T$. The order parameter $\Delta(z)$, chemical potential $\mu$, and the center of mass wavevector $q$ are to be obtained self-consistently together with the number equation, $N = \int d\br [ n_{\uparrow}(\br) + n_{\downarrow}(\br)]$, where $N$ is the total number of atoms and the density of atoms is given by
\be
\label{eq-nu}
n_{s} (z) = \frac{1}{2 N_x N_y} \sum_{n} |u_{s,n} (z) |^2 f(E_{n}) + |v_{s,n} (z) |^2 f(-E_{n}).
\ee
The set of solutions $\{\Delta_q, q, \mu \}$ of the self-consistent calculation with given $\lambda$, $h$ , and $\gamma$ can be obtained from the thermodynamic potential
\bea\nonumber
\Omega[\Delta_q, q, \mu]
&=&
-N_x N_y \int dz
\frac{|\Delta(z)|^2}{g_{1D}}
\\
&&+
\text{tr} \left[H_0 (z) \right]
+
\frac{1}{2}\sum_{n}
E_{n} f(E_{n})
\eea
by solving the set of non-linear equations self-consistently
\be
\frac{\partial \Omega }{ \partial \Delta_q }= 0,
\frac{\partial \Omega }{ \partial q }= 0,
N = -\frac{\partial \Omega }{ \partial \mu }.
\ee
We find that $\partial \Omega / \partial \Delta_q = 0$ and $N = - \partial \Omega / \partial \mu $ correspond to the order parameter equation (\ref{eq-op}) and the number equation (\ref{eq-nu}), respectively. Since, in general, the analytical expressions of eigenvalues $E_n$ are not available except for few simple cases, the partial derivatives need to be calculated numerically.

For the purpose of this paper, we assume the FFLO phase as the simplest case, that is, the Fulde-Ferrell (FF) single-plane-wave phase, $\Delta(z) = \Delta_{q} \exp [iqz]$, leading to
\be
u_{s,n} (z) = u_{s,\bk,q} \exp \left[ + i\left( \frac{q}{2} + k_z \right) z \right],
\ee
\be
v_{s,n} (z) = v_{s,\bk,q} \exp \left[ - i\left( \frac{q}{2} - k_z \right) z \right],
\ee
 so that the order parameter equation (\ref{eq-op}) reads
 \bea\nonumber
 \Delta(z) &=&
 -\frac{g_{1D} }{2N_x N_y} \sum_{\bk}
 \left[
 u_{\uparrow,\bk,q} v_{\downarrow,\bk,q}^\ast f(E_n)
 +
  u_{\downarrow,\bk,q} v_{\uparrow,\bk,q}^\ast f(-E_n)
  \right]
 e^{iqz}
 \\
 &=& \Delta_q e^{iqz}.
 \eea
 Even though the single-particle dispersion $\epsilon_\bk$ along $x$, $y$, and $z$ directions has inversion symmetry, thus in principle, the FFLO phase can host a finite FFLO momentum wave-vector $\bq$ in all directions, we ignore the interaction across the 1D tubes due to the tight-confinement of the atomic motion along the $z$ direction, and take the simplest ansatz: the FFLO order parameter with a single-component $\bq = q \hat{z}$.

 Then the BdG Hamiltonian in the momentum space becomes
\be
H_{BdG} (\bk)
=
 \bmatrix
 \xi_{\bk} - h & \lambda_{\bk} & 0 & -\Delta_q \\
 \lambda_{\bk}^\ast &  \xi_{\bk} + h & \Delta_q & 0 \\
 0 & \Delta_q^\ast & -\xi_{-\bk} + h & -\lambda_{-\bk}^\ast \\
 -\Delta_q^\ast & 0 & -\lambda_{-\bk} & -\xi_{-\bk} - h
 \ematrix,
\ee
where $\xi_{\pm \bk} = \epsilon_{(q/2) \hat{z} \pm \bk} - \mu$ and $\lambda_{\pm \bk} = -i \lambda (q/2 \pm k_z)$.

To understand topological superfluidity, we notice that the BdG hamiltonian has the particle-hole symmetry, which means the system is invariant under the particle-hole operator $i\tau_xC$, where $\tau$ is the Pauli matrix acting on the particle-hole space and $C$ is the complex conjugation operator.
The topological phase transitions is to be determined by the non-trivial number, $\mathcal{I} = \text{sign}[\text{Pf}(A_{sk})] = \pm 1$, where Pf is the Pfaffian of the skew matrix $A_{sk}$~\cite{kitaev-2003}.
For example, the skew matrix of BdG Hamiltonian $H_{BdG}(\bk)$ can be obtained as $H_{BdG} (0) (i \tau_x C )$. The topological non-trivial phase corresponds to $\mathcal{I} = -1$, while $\mathcal{I}=+1$ corresponds to the topologically trivial phase.
The topological nature, identified by $\mathcal{I} = \pm 1$, is protected by the gap in the bulk quasi-particle excitations.
From the eigenvalues $E_n$ of $H_{BdG}(\bk)$ we find the condition for the gap opening in the topologically non-trivial phase
\be
\label{eq-13}
h > h_t = \sqrt{|\Delta_q|^2 + \left[\frac{(q/2)^2}{2m}-\mu \right]^2 - \lambda^2 (q/2)^2},
\ee
together with non-zero $\lambda$, $\Delta_q$ and the gapped bulk excitation~\cite{,qu-2013-07,zhang-2013}.
Here, $h_t$ is the critical Zeeman field at which the excitation gap at $\bk = 0$ closes and beyond which it reopens. This condition $h> h_t$ is a generalized version of the condition for BCS topological superfluids ($q=0$), $h^2 > |\Delta_0|^2 + \mu^2$.

\section{Results}
\begin{figure}[t]
\includegraphics[width = 1.0\linewidth]{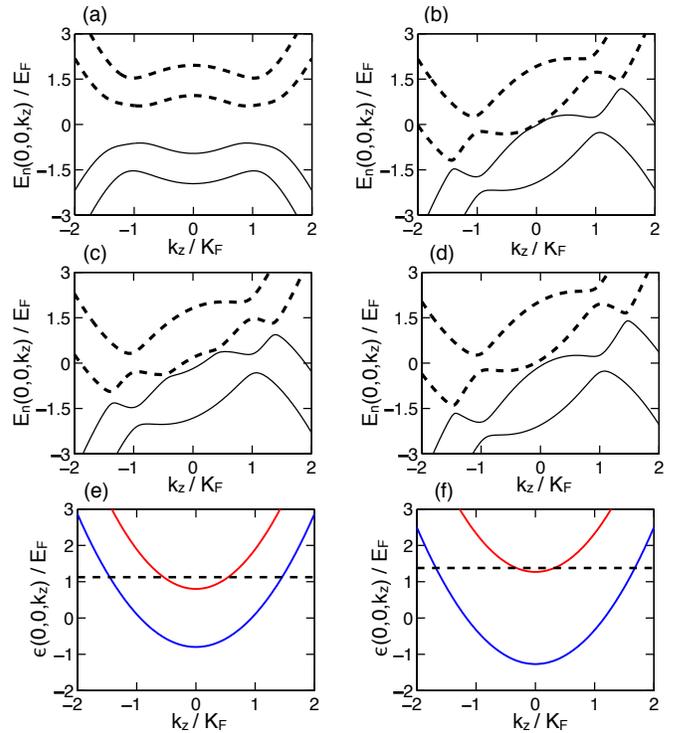}
\caption{
(Color online)
(a) The representative excitation spectrum of the quasi-particles and the quasi-holes as a function of $k_z$ in the $k_x = k_y = 0$ plane for the \textit{US} phase at $h / E_F = 0.5$ and $\lambda K_F = 0.2$. The spectrum is fully gapped and symmetric, as expected.
(b) The excitation spectrum at the critical Zeeman field $h = h_t =  0.93 E_F$ for the same SOC as (a).
The gap closing at $\bk = 0$ is consistent with the sign change in the Pfaffian at $h_t$, presumably indicating the TQPT within the FFLO phase.
(c) The excitation spectrum for the \textit{FFLO-1} phase with $q/K_F = 0.8$ at $h/E_F = 0.8<h_t$ and $\lambda K_F = 0.2$.
(d) The bulk spectrum for the \textit{FFLO-2} phase with $q/K_F=1.25$ at $h/E_F = 1.05 $ and $\lambda K_F = 0.2$.
(e) and (f) are the plots of the helicity bands $\epsilon_\bk = k_z^2/2m \pm \sqrt{ h^2 + (\lambda k_z)^2}$ for \textit{FFLO-1} and \textit{FFLO-2} phase at the same values of $h$ and $\lambda$ as in (c) and (d), respectively. The black dashed line indicates the chemical potential $\mu$ determined with the fixed atom number $N$.
}
\label{fig-3}
\end{figure}

We calculate the zero-temperature ($T$=0) superfluid phases of the spatially uniform quasi-one dimensional (1D) Fermi gas within mean-field theory in the 1D limit (small transverse hopping $t$), which captures most of the qualitative features of the phase diagram.
Fig.~\ref{fig-1} shows a mean-field phase diagram in the Zeeman field $h$ versus spin-orbit coupling $\lambda$ plane computed with a fixed transverse hopping $t/E_F=0.01$, corresponding to fixed optical lattice intensity, and the dimensionless interaction parameter $\gamma = -2$.
We find that the uniform superfluid phase (\textit{US}), which is characterized by the order parameter $\Delta_q$ with the zero center of mass momentum $q = 0$ Cooper pairs, is stable against non-uniform FFLO phases for the weak Zeeman field ($h<h_c$) and spin-orbit coupling, with the phase boundary indicated by the solid line.
The FFLO phase results from the condensate of the Cooper pairs with finite center of mass momentum ($q \neq 0$) and the Zeeman field facilitates in creating such pairs with finite total momentum by shifting the up-spin and the down-spin bands relative to each other.
Notice that in the absence of SOC with increasing Zeeman field further ($h/E_F > 2$ for $\gamma = -2$) the FFLO phase enters into the normal state via second order phase transition, while the finite SOC prevents the phase transition to normal state.
In the presence of the SOC, the pairing mechanism for the FFLO superfluid phase is different from that without SOC because it can be due to the pairing of the states 
within, also between, the two helicity bands. 
Note that in addition to finite $q$ pairing between the two helicity bands, the Zeeman field and (weak) SOC induced momentum distributions of the states $n_{\bk,s}$ within the same helicity band 
may also lead to the pairing of states $|\bk + q/2 \hat{z},\uparrow\rangle$ and $|-\bk + q/2 \hat{z},\downarrow\rangle$ to form an FFLO phase. 
In general, however, SOC disfavors creation of the FFLO phase (see the upturn of the US to FFLO phase boundary in Fig.~\ref{fig-1}).
The large enough SOC relative to the Zeeman field make $ n_{\bk \uparrow} - n_{\bk\downarrow}$ negligible thus lowering the energy of the uniform superfluid state.
Above a threshold value of the SOC the phase transition between the \textit{US} and the FFLO phase is removed altogether and with further increase of the SOC, with increasing values of the Zeeman field, the \textit{US} phase has an instability to a uniform \textit{topological} superfluid (\textit{UTS}) state carrying Majorana fermion excitations at the tube ends.
We find that the phase transition between the \textit{US} and the \textit{UTS} is second order and the energy gap vanishes at the critical point $h_t  = \sqrt{|\Delta_0|^2 + \mu^2}$  (dashed line).
Above the critical point ($h > h_t$),
Fig~\ref{fig-2}(b) shows that  the energy gap in the \textit{UTS} phase is re-opened and finite everywhere in the momentum space, thus protecting the zero energy Majorana fermions at the tube ends.
At the phase boundary ($h = h_t$) the so-called Pfaffian topological invariant, given in Eq.~(\ref{eq-13}), changes sign. This is only possible if the corresponding energy gap vanishes at $\bk=0$ (for example, FIG.~\ref{fig-3}(b) for the FFLO phase).
Interestingly, in the regime where Zeeman field $h$ is strong enough ($h>h_c$) but with weak spin-orbit coupling $\lambda$, we find another topological phase transition within the FFLO phase, characterized by this sign change of the Pfaffian topological invariant: TQPT between \textit{FFLO-1} and \textit{FFLO-2}.
In contrast to the bulk gapped topological uniform superfluid (\textit{UTS}), FIG.~\ref{fig-2}(a) shows that the finite gap in the \textit{US} phase closes at the critical field $h=h_c$  and remains gapless for the Zeeman field above $h_c$, resulting in the gapless non-uniform FFLO phases.
In addition to the phase transitions along the Zeeman field, we can see how the superfluid phases depend on the SOC $\lambda$ from the FIG.~\ref{fig-2}(c) and (d) showing the order parameter $\Delta_q$ and the chemical potential $\mu$ as a function of $\lambda$ during the \textit{FFLO-1}--\textit{US} and \textit{FFLO-2}--\textit{US} phase transitions (open circle marks $h_c$) as well as the \textit{US}--\textit{UTS} topological phase transitions (open square marks $h_t$). The SOC $\lambda$ enhances the order parameter $\Delta_q$ in the \textit{FFLO-1} phase, while the order parameter in the \textit{FFLO-2} phase decreases with increasing $\lambda$. The chemical potential $\mu$ decreases with increasing $\lambda$. Particularly, it shows the prominent change in the slope at the \textit{US}--\textit{UTS} topological phase transition.

Now let us consider the momentum-resolved topological properties of the non-uniform FFLO phases.
Fig.~\ref{fig-3}(a) shows the excitation spectrum of quasi-particle (dashed lines) and quasi-hole (solid lines) bands at $h <h_c$ and for finite $\lambda$ as a function of $k_z$. We see that the spectrum in the \textit{US} phase is fully gapped and symmetric with respect to $+k_z$ and $-k_z$. The finite gap between the quasi-particle and the quasi-hole bands is due to the uniform superfluid order parameter, while the Zeeman field induces a splitting \textit{within} the particle and the hole bands.
At strong Zeeman field $h > h_c$, the \textit{US} phase
becomes energetically unstable against the Fulde-Ferrell (FF) superfluid phase. 
Fig.~\ref{fig-3}(b) shows that the excitation gap at $\bk = 0$ closes at the critical point $h=h_t$ associated with the sign change in the non-trivial Pfaffian topological index, indicating the \textit{FFLO-1}--\textit{FFLO-2} transition.
In Figs.~\ref{fig-3}(c) and (d) we plot the quasi-particle and quasi-hole excitation bands of these topologically distinct FFLO phases, \textit{FFLO-1} ($h_c < h<h_t$) and \textit{FFLO-2} ($h > h_t$) for the same spin-orbit coupling as in Fig.~\ref{fig-3}(a).
We can see that the excitation bands for both topologically distinct FFLO phases are asymmetric with respect to $+k_z$ and $-k_z$ and gapless as we have shown in FIG.~\ref{fig-2}(a).
We understand from FIG.~\ref{fig-3}(e) that the \textit{FFLO-1} phase is topologically trivial since the chemical potential $\mu$ crosses both helicity bands, while as shown in FIG.~\ref{fig-3}(f) the chemical potential in the same ($h$, $\lambda$) parameter region for the \textit{FFLO-2} phase is located very close to the bottom of the upper helicity band. 
In contrast to the bulk energy gap in the \textit{UTS} phase, however, the \textit{FFLO-2} phase is a gapless phase with a non-trivial Pfaffian topological index. A full characterization of this phase should be interesting and we keep this for future study.

%

\begin{figure}[t]
\includegraphics[width = 1\linewidth]{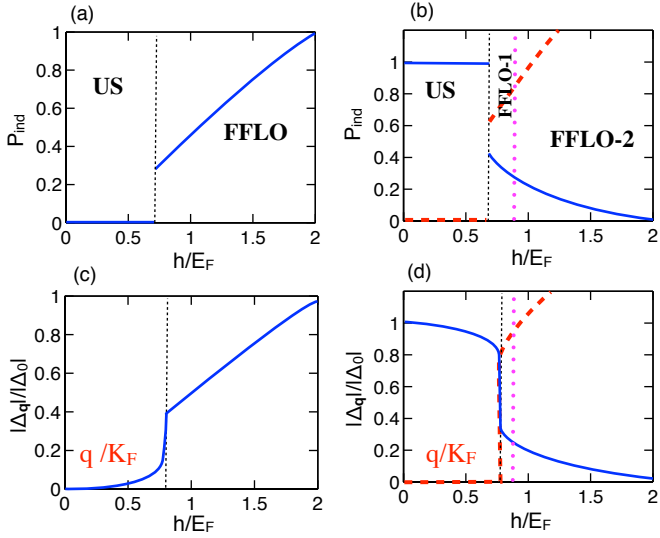}
\caption{
(Color online) (a) Induced polarization $P_{ind} = (N_\uparrow - N_\downarrow)/N$ as a function of Zeeman field $h$ for $\lambda = 0$. The \textit{US} phase is unpolarized ($P_{ind} = 0$), while the normal phase ($h/E_F>2$) is fully polarized ($P_{ind} = 1$). $P_{ind}$ in the FFLO phase increases with increasing $h$.
(b) $P_{ind}$ as a function of $h$ for the finite SOC $\lambda K_F = 0.2$. In contrast to the $\lambda = 0$ case, the \textit{US} phase is polarized for the finite Zeeman field. The purple dotted line indicates the TQPT between \textit{FFLO-1} and \textit{FFLO-2} phases.
(c) The order parameter $|\Delta_q|$ (blue solid lines) and the center of mass momentum $q$ (red dashed lines) as a function of $h$ without spin-orbit coupling $\lambda = 0$. With increasing $h$, in the FFLO phase the order parameter $|\Delta_q|$, normalized by $|\Delta_0|$, decreases, while the center of mass momentum $q$ increases. For the Zeeman field below the critical value $h < h_c$, the order parameter stays constant, $\Delta_q = \Delta_0$.
(d) $\Delta_q$ and $q$ as a function of $h$ in the presence of SOC ($\lambda K_F = 0.2$). In contrast to the \textit{US} phase in (c), with increasing $h$, the $\Delta_q$ slightly decreases with the center of mass momentum $q$ staying at zero.
Here, we used the interaction parameter $\gamma = -2$ and tunneling amplitude $ t/E_F = 0.01$ as in the Fig.~\ref{fig-1}.
}
\label{fig-4}
\end{figure}

Zeeman field $h$ induces the polarization $P_{ind} = (N_\uparrow - N_\downarrow)/N$, or spin-polarization, by increasing the effective chemical potential for spin-up, $\mu_\uparrow = \mu + h$, and reducing that for spin-down, $\mu_\downarrow = \mu - h$. In a uniform superfluid without spin-orbit coupling, one needs to apply the Zeeman field over a threshold value to induce a non-zero spin polarization. However, in the presence of a finite spin-orbit coupling any non-zero Zeeman field can induce a spin polarization even without requiring a threshold due to the mixing of the spin states in the same band.
In Fig.~\ref{fig-4}(a) and (b), we show how the Zeeman field $h$ induces the polarization $P_{ind}$ in the superfluid phases in the case of zero spin-orbit coupling ($\lambda = 0$) and finite spin-orbit coupling ($\lambda K_F = 0.2$) with $\gamma = -2$, respectively. Without spin-orbit coupling (Fig.~\ref{fig-4}(a)), the \textit{US} phase is population balanced ($N_\uparrow = N_\downarrow$), or has zero spin-polarization, while the Zeeman field induces a finite spin-polarization in the FFLO phase above $h_c$.
With increasing $h$ further, the system becomes fully polarized ($P_{ind} = 1$), and enters into the normal ($N$) phase via a second order phase transition.
In contrast, the spin mixing due to the spin-orbit coupling always keeps $P_{ind} < 1$  and the system is always in a spin-polarized superfluid phase (Fig.~\ref{fig-4}(b)).
Rather than the phase transition to normal state, the \textit{FFLO-1} in the weak spin-orbit coupling regime becomes unstable against the topologically distinct FFLO phase (\textit{FFLO-2}) at $h = h_t$, which is indicated by the purple dotted line.
In the very high Zeeman field regime ($h/E_F > 2$ for $\lambda = 0$), corresponding to the normal state, the system is still not fully polarized. Thus it is capable of remaining in a superfluid phase, albeit with extremely small pairing amplitude $|\Delta_q|$ and with finite center of mass momentum $q$.
Fig.~\ref{fig-4} (c) and (d) show the order parameter $\Delta_q$ (blue solid lines) and the pairing momentum $q$ (red dashed lines) as a function of Zeeman field $h$ for $\lambda = 0$ and $\lambda K_F = 0.2$, respectively.
We see from Fig.~\ref{fig-4}(c) that the order parameter in the \textit{US} phase with zero SOC remains constant with $P_{ind} = 0$ and $q = 0$ until $h$ reaches the phase boundary $h_c$, where the system begins polarized $P_{ind} >0$, leading to the phase transition from \textit{US} to FFLO phase.
The center of mass momentum $q$ increases with increasing Zeeman field $h$, mimicking the behavior of $P_{ind}$ in the FFLO phase.
In the presence of the SOC (Fig.~\ref{fig-4}(d), the order parameter $\Delta_q$ in the \textit{US} phase is slightly decreasing with increasing $h$, while $q$ remains zero. In contrast to the case in Fig.~\ref{fig-4}(c), with increasing $h$ and for a finite $\lambda$, the center of mass momentum $q$ in the FFLO phases increases, while the order parameter $\Delta_q$ decreases and asymptotically becomes zero.
It is manifest from the featureless behaviors of $\Delta_q$ and $q$ in Fig.~\ref{fig-4}(d)
that the topological phase transition at $h= h_t$ is not associated with any global symmetry breaking of the system,
\begin{figure}[t]
\includegraphics[width = 0.8\linewidth]{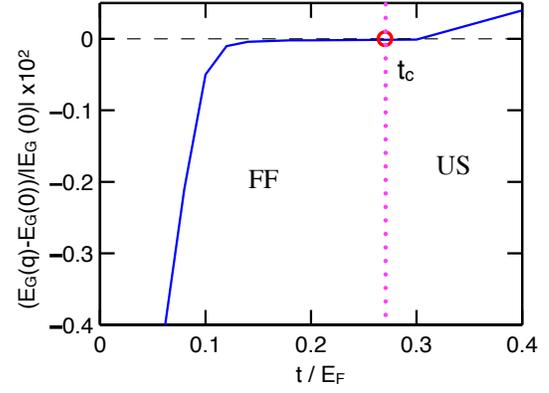}
\caption{
(Color online) The difference in the ground state energy (($E_G(q)-E_G(0)$)) between the FFLO superfluid phase and the uniform superfluid phase as a function of the transverse hopping $t$. We have used a fixed Zeeman field $h/E_F = 1$ and a fixed spin-orbit coupling strength $\lambda K_F = 0.2$ with the interaction parameter $\gamma = -2$. With increasing $t$ (the system becoming more and more 3D-like) the ground state energy of the FFLO phase becomes larger and eventually becomes bigger than the ground state energy of the uniform phase. For the transverse hopping above the value $t_c/E_F = 0.28$ (red open circle), the ground state of the system crosses over to become the uniform superfluid phase. The purple dotted line indicates the phase boundary between the FFLO and uniform superfluid phases.}
\label{fig-5}
\end{figure}

Now, we consider the nature of the FFLO superfluids and their topological features during the dimensional crossover from the 1D to 3D regime by increasing the transverse hopping $t$, associated with the optical lattice intensity.
Fig.~\ref{fig-5} presents the difference in the ground state energy $(E_G(q) - E_G(0))/|E_G(0)|$, where $E_G(q)$ is the ground state energy with the center of mass momentum $q$ at $T=0$. The latter can be obtained from the Helmholtz Free energy $F = E-TS = \Omega + \mu N$ through a limit of $T \to 0$
\be
E_G(q) =
-N_x N_y L_z\frac{|\Delta_q|^2}{g_{1D}} + \frac{1}{2} \sum_n E_n \theta(-E_n)
+ \sum_{\bk} \xi_{-\bk} + \mu N,
\ee
where $\theta(x)$ is a Heaviside step function~\cite{zeng-2013}.
With the self-consistent solutions $\{ \Delta_q, \mu, q\}$, we evaluate the difference in the zero-temperature ground state energy between FFLO phase, $E_G(q)$, and a uniform superfluid phase, $E_G(0)$, as function of the transverse hopping $t$ at fixed values of Zeeman field ($h/E_F= 1$) and SOC strength ($\lambda K_F = 0.2$) with $\gamma = -2$, corresponding to the gapless topological FFLO phase (\textit{FFLO-2}) for $t= 0.01E_F$.
With increasing the transverse hopping $t$, the quasi-1D system becomes more and more 3D-like. As shown in Fig.~\ref{fig-4}, the ground state energy of the FFLO superfluid phase increases and eventually becomes bigger than the ground state energy of the uniform superfluid phase for the transverse hopping above the critical value $t_c$ ($ = 0.28 E_F $ for $h/E_F = 1$ and $\lambda K_F = 0.2$ with $\gamma = -2$), leading to the phase transition from the quasi-1D FFLO superfluid (\textit{FFLO-2}) to the uniform superfluid (\textit{UTS}).

\section{Conclusions}
In conclusion, we have investigated theoretically the properties of quasi-one dimensional Fermi superfluidity under a non-Abelian synthetic gauge field and Zeeman field.
We have shown that the zero-temperature ($T=0$) phase diagram of quasi-1D Fermi gas (``tubes'') in a two-dimensional lattice with small transverse hopping $t$ (the quasi-1D regime) exhibits the topologically distinct single-plane-wave FFLO phases (\textit{FFLO-1}, \textit{FFLO-2}) in the presence of small Rashba type spin-orbit coupling $\lambda$ and an effective Zeeman splitting $h$ at $h > h_c$.
The uniform superfluid phase (\textit{US}) with the zero center of mass momentum ($q=0$) Cooper pairs takes over the phase diagram even for high values of the Zeeman field for larger values of the spin-orbit coupling. For even higher spin-orbit coupling (see Fig.~\ref{fig-1}), a high Zeeman field induces the topological phase transition within the uniform superfluid phase to a \textit{topological} uniform superfluid phase with Majorana fermions at the ends of the 1D tubes.
In 1D regime, corresponding to small $t$, for small values of the SOC, we have found that a topological phase transition can take place within the FFLO phase at which the Pfaffian topological index changes sign but the system remains gapless. The principal result of this paper is that, as the system crosses over from 3D to quasi-1D with decreasing transverse hopping, the 3D uniform superfluid phase in a Zeeman field crosses over to the FFLO phase for small values of the SOC but to the topological uniform superfluid state with Majorana fermion excitations for larger values of the SOC. Therefore, the SOC separates the two important unconventional superfluid phases in very different regimes of the parameter space, which should be useful for experimental realizations of these phases in the laboratory.
We have also investigated the crossover physics from  1D to 3D regime by increasing the lattice intensity, corresponding to increasing the transverse hopping $t$ in the single-particle dispersion $\epsilon_\bk$. As anticipated in the phase diagram of 3D Fermi gas, the FFLO phase is diminished with increasing $t$ with $h$ and $\lambda$ fixed, and the uniform superfluid phase becomes favorable energetically.

\begin{acknowledgments}
We thank Chunlei Qu for valuable comments.
This work is supported by AFOSR (FA9550-13-1-0045), 
NSF-PHY (1104546), 
NSF-PHY (1104527),
and ARO (W911NF-12-1-0334).
\end{acknowledgments}



\end{document}